\documentstyle[aps,epsfig]{revtex}
\begin{document}
\draft{}
\title{First observation of 
$\phi(1020)\to\pi^0\pi^0\gamma$ decay}

\author{
M.N.Achasov, V.M.Aulchenko, A.V.Berdyugin, A.V.Bozhenok, A.D.Bukin,
D.A.Bukin, S.V.Burdin, T.V.Dimova, S.I.Dolinsky, V.P.Druzhinin,
M.S.Dubrovin, I.A.Gaponenko, V.B.Golubev, 
V.N.Ivanchenko \thanks{email:V.N.Ivanchenko@inp.nsk.su},
I.A.Koop,
A.A.Korol, S.V.Koshuba, E.V.Pakhtusova, E.A.Perevedentsev, A.A.Salnikov,
S.I.Serednyakov, V.V.Shary, Yu.M.Shatunov, V.A.Sidorov, Z.K.Silagadze,
A.N.Skrinsky, Yu.V.Usov, Yu.S.Velikzhanin
}

\address{Budker Institute of Nuclear Physics, Novosibirsk, 630090, Russia}

\date{\today}

\maketitle

\begin{abstract}
In the SND experiment at VEPP-2M $e^+e^-$ collider  the
$\phi(1020)\to\pi^0\pi^0\gamma$ decay was studied. Its branching ratio
$B(\phi \to \pi^0\pi^0\gamma )=(1.14 \pm 0.10 \pm 0.12) \cdot 10^{-4}$ 
was measured. 
It was shown, that $f_0(980)\gamma$ mechanism dominates in this
decay. Corresponding branching ratio 
$B(\phi \to f_0\gamma ) = (3.42\pm0.30\pm0.36) \cdot 10^{-4}$
was obtained. 
\end{abstract}


\vspace*{3mm}


\nopagebreak
\twocolumn

\section{Introduction}

First search for $\phi\to\pi^0\pi^0\gamma$ decay was carried out
with the ND detector at VEPP-2M $e^+e^-$ 
collider in 1987 \cite{IVN1,REP}. 
In this early experiment the upper limit
$B(\phi\to\pi^o\pi^o\gamma) < 10^{-3}$ was placed.
As it was shown later by Achasov \cite{IVN2}, study of this decay can provide a
unique information on the structure of the light scalar $f_0(980)$ meson.
Subsequent studies \cite{IVN3,IVN5,IVN6,IVN7,IVN8,IVN9}
proved this idea. In these works different
models of the  $f_0(980)$-meson structure were considered. The most
popular were 2-quark model \cite{IVN12}, 4-quark MIT-bag model \cite{IVN13}, 
and $K\overline{K}$ molecular model \cite{IVN6}.

 In 1995 new Spherical Nonmagnetic Detector
(SND), having better hermeticity, granularity,
energy and spatial resolution than previous ND detector,
started operation at VEPP-2M. First indications of the process
\begin{equation}
{e^+e^-\to\phi\to\pi^0\pi^0\gamma}
\label{EQ0}
\end{equation}
were seen by SND in 1997 \cite{FIRST}.
This preliminary
result was based on analysis of half of the $\phi$-meson
data recorded in 1996 -- 1997 \cite{FI96}.
Present work is based on the analysis of
full data sample. The $\phi\to\pi^0\pi^0\gamma$ decay was observed
for the first time and its branching ratio was measured.

\section{Detector and experiment}

The  SND detector \cite{SND} is designed for the 
study of $e^+e^-$-annihilation at center of mass energy about $1~GeV$.
Its main part is a three layer electromagnetic calorimeter,
consisted of 1630 NaI(Tl) crystals \cite{CAL}.
The energy resolution of the calorimeter
for photons can be described as 
$\sigma_E/E=4.2\%/ \sqrt [{4\;\;}] {E(GeV)} $ \cite{CAL1,CAL2},
angular resolution is close to $1.5$ degrees.
The 1996 experiment consisted of 7 independent runs \cite{FI96}.
In each run the data were recorded at 14 different beam
energies in the region $2E_0=(980-1050)~MeV$ covering
the peak and close vicinity of the $\phi$-resonance.
The energy spread of the beam energy was equal to 0.2~MeV.
The total integrated luminosity of $3.9~pb^{-1}$ was collected,
corresponding to about $8\cdot10^6$ 
$\phi$ mesons produced. The luminosity determination was based on
events of Bhabha scattering and two-photon annihilation detected in
the calorimeter. Accuracy of normalization is better than $3~\%$.

\section{Data analysis}

Main resonant background to the decay  (\ref{EQ0})
comes from the decay
\begin{equation}
{e^+e^-\to\phi\to\eta\gamma\to3\pi^0\gamma}
\label{EQ1}
\end{equation}
due to the merging of
photons and/or loss of photons through the openings in the
calorimeter.
The main source of non-resonant background is a process
\begin{equation}
{e^+e^-\to\omega\pi^0\to\pi^0\pi^0\gamma.}
\label{EQ3}
\end{equation}
The background from the $\phi\to\rho\pi^0\to\pi^0\pi^0\gamma$
decay is small \cite{IVN5,IVN11}, nevertheless its
amplitude was included into simulation of the process (\ref{EQ3}).
The background from the QED $5\gamma$-annihilation process was
estimated and found to be negligible.

In order to suppress the contribution of background 
the events were selected with 5 photons 
emitted at angles of more than 27 
degrees with respect to the beam and without charged particles.
Standard SND cuts 
 on energy-momentum balance   in an event 
and on photon identification function were used:

\begin{itemize}
\item  
$0.8 < E_{dep}/2E_0 < 1.1$ , where $E_0$ is the beam energy,
$E_{dep}$ - the total energy deposition in the calorimeter;
\item  $p/2E_0 < 0.15$, where $p$ -- is a total momentum of all photons
in the event;
\item $\chi^2 < 20$ -- event kinematics is
consistent with $\pi^0\pi^0\gamma$ hypothesis;  
\item  $\zeta < 0$;
\end{itemize}

The  $\zeta$ parameter was constructed on the basis of the logarithmic
likelihood function, describing the probability for observed
transverse profile of electromagnetic shower in the calorimeter to be
generated by a single photon \cite{FIRST,GKL}. 
This parameter 
%(Fig.\ref{fig1})
facilitates efficient
separation of events with well isolated photons from those
with merged photons or with showers, produced by  $K_L$ mesons.

\begin{figure}[b!] % fig 1
\centerline{\epsfig{file=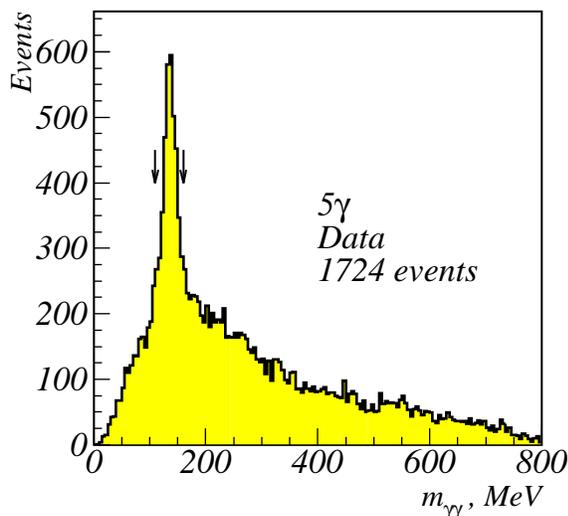}}
\vspace{1pt}
\caption{Invariant mass spectrum of photon pairs for selected 
5-gamma events.}
\label{fig2}
\end{figure}

The $\chi^2$ is a  kinematic fit parameter, describing the degree of
energy-momentum conservation in the event
with additional
requirement that two $\pi^0$ mesons are present. During this fit all
possible combinations of photons 
(Fig.\ref{fig2})
in the event were checked in a search for
invariant masses $m_1$ and  $m_2$, satisfying the condition:
$\sqrt{(m_1-m_{\pi})^2+(m_2-m_{\pi})^2}<25~MeV/c^2.$

The  $\phi\to K_S K_L \to\pi^0\pi^0K_L$ decay
can contribute due to nuclear interactions
of $K_L$ mesons in the material of the calorimeter but after described cuts 
  it is not seen at present level of statistics.

\begin{figure}[b!] % fig 4
\centerline{\epsfig{file=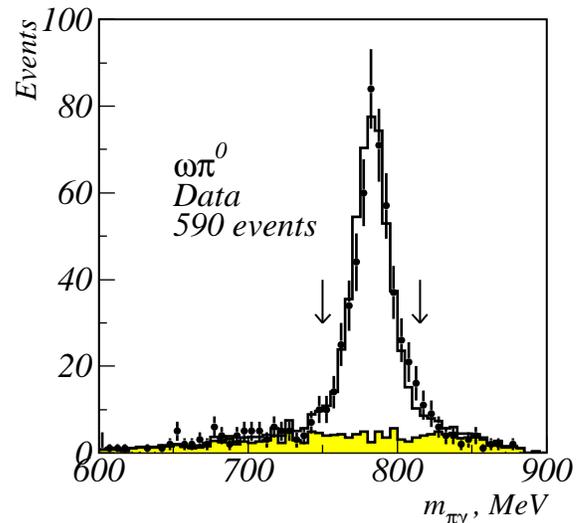}}
\vspace{1pt}
\caption{Distribution over $\pi^0\gamma$ invariant mass for
$m_{\pi\pi}<700~MeV$. 
Points -- data, histogram -- simulation, shaded histogram --
sum of simulated contributions from $\phi\to\eta\gamma$ and 
$\phi\to\pi^0\pi^0\gamma$ decays,
arrows -- selection of the  $\omega\pi^0$ class.}
\label{fig4}
\end{figure}

In the energy region of this experiment the invariant mass of the
pion pair in the process (\ref{EQ3}) is less than $700~MeV$. In the events
satisfying this condition clear $\omega$(782) peak is seen
in $m_{\pi\gamma}$ distribution (Fig.\ref{fig4}), proving
the domination of the process (\ref{EQ3}) in this kinematic region.
The $m_{\pi\gamma}$ parameter was defined as an invariant mass of the
recoil photon and one of $\pi^0$ mesons, closest to the $\omega$-meson mass.
499 found events with
$750~MeV < m_{\pi\gamma} < 815~MeV$ were assigned to $\omega\pi^0$
class, while 189 events with $m_{\pi\gamma}$ outside this interval and 
$m_{\pi\pi}>700~MeV$ were assigned to  $\pi^0\pi^0\gamma$ class.
Subtracting calculated contribution of the process (\ref{EQ1}) and 
using estimated probabilities of  events misidentification for
the processes (\ref{EQ0}) and (\ref{EQ3}) the number
of the events of the process  (\ref{EQ3}) in the $\omega\pi^0$
class was estimated to be equal to 449.
The corresponding number of events of the decay (\ref{EQ0})
in the   $\pi^0\pi^0\gamma$ class is 164.
The background from the process (\ref{EQ3}) was estimated using events
of the $\omega\pi^0$ class. No additional knowledge of
the actual production cross section of this process was necessary.

\begin{figure}[b!] % fig 5
\centerline{\epsfig{file=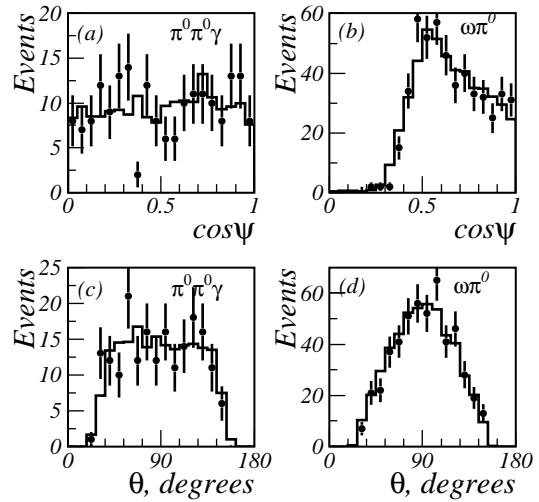}}
\vspace{1pt}
\caption{a, b -- cosine of $\psi$,
the angle between directions of $\pi^0$ and recoil $\gamma$
in the rest frame of $\pi^0\pi^0$ system;
c, d --  distributions in $\theta$, angle of the
recoil $\gamma$ with respect to the beam.
Points -- data, histogram -- simulation.}
\label{fig5}
\end{figure}

Then, for the events of the $\pi^0\pi^0\gamma$ and $\omega\pi^0$
classes the comparison of experimental and simulated
distributions in  $\psi$ and $\theta$ angles was done.
Here $\psi$ is an angle of the recoil photon with respect to pion
direction in the $\pi^0\pi^0$ center of mass reference frame, 
$\theta$ is an angle between recoil photon and the beam.
The distribution in  $\theta$ for  $\pi^0\pi^0\gamma$
events with pion pair in a scalar state must be proportional to
$1+cos^2\theta$ and uniform in $cos\psi$.
The comparison (Fig.\ref{fig5}a,c) shows, 
that in $\pi^0\pi^0\gamma$ class of events 
pions are actually produced in scalar state. 
On the contrary, the experimental
events of the $\omega\pi^0$ class (Fig.\ref{fig5}b,d)
well match the hypothesis of the intermediate $\omega\pi^0$ state
with quite different $\psi$ distribution.

\begin{figure}[b!] % fig 6
\centerline{\epsfig{file=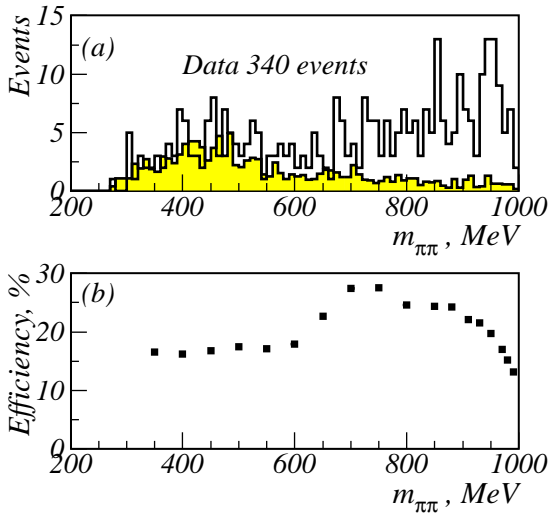}}
\vspace{1pt}
\caption{a -- invariant mass distribution
of $\pi^0\pi^0$ pairs for selected $\pi^0\pi^0\gamma$ events
without acceptance corrections. 
Histogram -- data, shaded histogram -- estimated background
contribution from $e^+e^-\to\omega\pi^0$ and
$\phi\to\eta\gamma$;
b -- detection efficiency for  $\pi^0\pi^0\gamma$ events.}
\label{fig6}
\end{figure}

The $\pi^0\pi^0$ invariant mass distribution for the events with 
$m_{\pi\gamma}$ outside the $750~MeV < m_{\pi\gamma} < 815~MeV$ interval
(Fig.\ref{fig6}a) 
shows significant excess over background
at large $m_{\pi\pi}$. At $m_{\pi\pi}<600~MeV$
the sum of
background contributions dominates.
Detection efficiency (Fig.\ref{fig6}b)
for the process (\ref{EQ0}) was determined using simulation 
of the process   $\phi\to S \gamma \to \pi^0 \pi^0\gamma$, where $S$
is a scalar state
with a mass ranging from 300 to $1000~MeV$ and zero width. 
In addition, this simulation provided information on $\pi^0\pi^0$
invariant mass resolution and event misidentification probability
as a function of $m_{\pi\pi}$.

After background subtraction,  correction for detection
efficiency, and correction for misidentification
 the mass spectrum was obtained (Table 1). 
For masses in the  $(600-850)~MeV$ interval the invariant mass resolution
is equal to $12~MeV$, so the $20~MeV$ bin size was chosen. At higher masses
the resolution improves, reaching $7.5~MeV$ at $950~MeV$, thus the bin size
of $10~MeV$ was used for higher masses.

To check accuracy of the detection efficiency, obtained by simulation,
events of the process (\ref{EQ1}) with 7 photons
in the final state and reconstructed  3 $\pi^0$ mesons were analyzed
\cite{ETG}. All other
selection criteria were the same as in the $\pi^0\pi^0\gamma$ analysis.
The number of observed events of the process (\ref{EQ1}) 
together with the PDG Table value for
$B(\phi\to\eta\gamma)=(1.26\pm0.06)~\%$ \cite{PDG}
and known total integrated luminosity 
provided independent efficiency estimation, based on experimental data.
Obtained 10\%
difference between experimental and simulated detection efficiencies was
used as a correction to the $\pi^0\pi^0\gamma$ detection efficiency,
obtained by simulation.
Such an approach minimizes systematic errors corresponding
to inaccurate simulation of tails of distributions over parameters, used for
event selection.

\section{Results}

Summing data from the Table 1, one can estimate branching ratio
of the decay (\ref{EQ0}) for the mass range $m_{\pi\pi}>700~MeV$
\begin{equation}
{B(\phi\to\pi^0\pi^0\gamma)=(1.00\pm 0.07\pm0.12)\cdot10^{-4},}
\label{EQ5}
\end{equation}
and for $m_{\pi\pi}>900~MeV$
\begin{equation}
{B(\phi\to\pi^0\pi^0\gamma)=(0.50\pm 0.06\pm0.06)\cdot10^{-4}.}
\label{EQ6}
\end{equation}
Here the first error is statistical and the second is systematic,
which was estimated to be close to $12~\%$. The systematic error
is mainly determined by the following contributions:

\begin{itemize}
\item the background subtraction error, which decreases almost linearly
with the increase of invariant mass  $m_{\pi\pi}$ and is  $5~\%$ on average;
\item error in the detection efficiency estimation, which increases with
 $m_{\pi\pi}$ and is equal to $8~\%$ on average;
 \item systematic error in $B(\phi\to\eta\gamma)$ is equal to $5~\%$.
 \end{itemize}

 The $m_{\pi\pi}$ invariant mass spectrum (Fig.\ref{fig8}) was fitted
according to Refs.\cite{IVN2}  
and further used for simulation of the decay (\ref{EQ0}). 
As a result the detection efficiency of the process (\ref{EQ0}) was
 estimated
 (14.72~\%) for invariant masses within the $(600-1000)~MeV$ interval
(Fig.\ref{fig6}a).
 This efficiency was used in the fitting of the $\phi$-resonance excitation
 curve. The visible cross section in each energy point $\sigma_{vis}(s)$
was described as a sum of the processes
 (\ref{EQ0}), (\ref{EQ1}), 
and (\ref{EQ3}) 
with efficiency and  radiative 
corrections taken into account for each process. 
 The background cross section due to the process (\ref{EQ3}) 
was estimated by fitting the visible
 cross section of the events of the $\omega\pi^0$ type
with linear function. The background from the
 process (\ref{EQ1}) was obtained from simulation. 
The only free parameter of the fit was the
$\phi\to\pi^0\pi^0\gamma$ branching ratio, all other
$\phi-$meson parameters were taken from the PDG data \cite{PDG}.
 As a result  (Fig.\ref{fig7}), the following value was obtained:
 \begin{equation}
 {
 B(\phi\to\pi^0\pi^0\gamma)=(1.14\pm 0.10\pm0.12)\cdot10^{-4},
 }
 \label{EQ7}
 \end{equation}
 which, in contrast with (\ref{EQ5}) and (\ref{EQ6}), 
is valid for the whole mass
 spectrum. In the systematic error estimation the following considerations
 were taken into account. 
In comparison with the results (\ref{EQ5}) and (\ref{EQ6}) the
 accuracy of normalization  $3~\%$ and efficiency estimation
 $5~\%$ are higher here, the background subtraction error $5~\%$  is
 the same, but an additional systematic error  $6~\%$ exists, due to
 uncertainty in extrapolation of the invariant mass spectrum into the region
 $m_{\pi\pi}<600~MeV$.
 Smaller systematic uncertainty has a ratio of branching ratios:
 \begin{equation}
 {\frac{B(\phi\to\pi^0\pi^0\gamma)}{B(\phi\to\eta\gamma)}
 =(0.90\pm 0.08\pm0.07)\cdot10^{-2},
 }
 \label{EQ8}
 \end{equation}
 Assuming that the process (\ref{EQ0}) is fully determined by  $f_0\gamma$
 mechanism, using the relation $B(f_0\to\pi^+\pi^-)=2B(f_0\to\pi^0\pi^0)$,
 and neglecting the decay $\phi\to KK\gamma$ 
\cite{IVN2}, we can obtain from (\ref{EQ7})
 \begin{equation}
 {
 B(\phi \to f_0(980) \gamma) = (3.42 \pm 0.30\pm0.36)\cdot 10^{-4}.
 }
 \label{EQA}
 \end{equation}
To monitor stability of experimental results, the number of
$\pi^0\pi^0\gamma$ events was checked separately in all
7 experimental runs. The results for all runs agree well within
statistical uncertainty. Another important proof of
validity of the results is the study of the process (\ref{EQ1}) \cite{ETG},
which properties are very close to the decay under study.
The methods of analysis are also very close, and the measured value
of the decay branching ratio (\ref{EQ1}) agrees with the PDG table value
\cite{PDG}.

\section{Discussion}

\begin{figure}[b!] % fig 8
\centerline{\epsfig{file=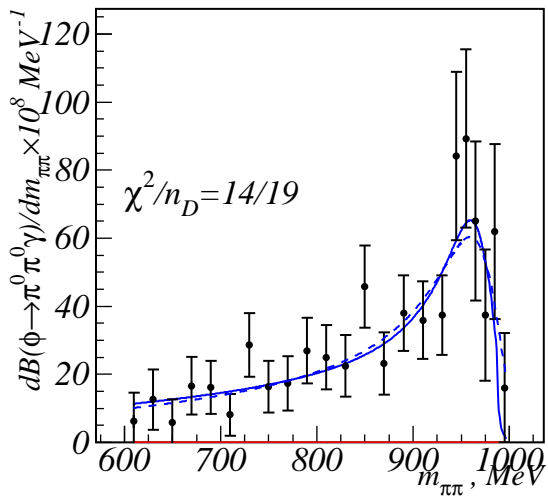}}
\vspace{1pt}
\caption{The measured $\pi^0\pi^0$ invariant mass spectrum.
Background is subtracted and efficiency corrections applied. 
Points -- data, solid line -- the result of the ``broad resonance'' fit,
dashed line -- the result of the ``narrow resonance'' fit.}
\label{fig8}
\end{figure}

\begin{figure}[b!] % fig 7
\centerline{\epsfig{file=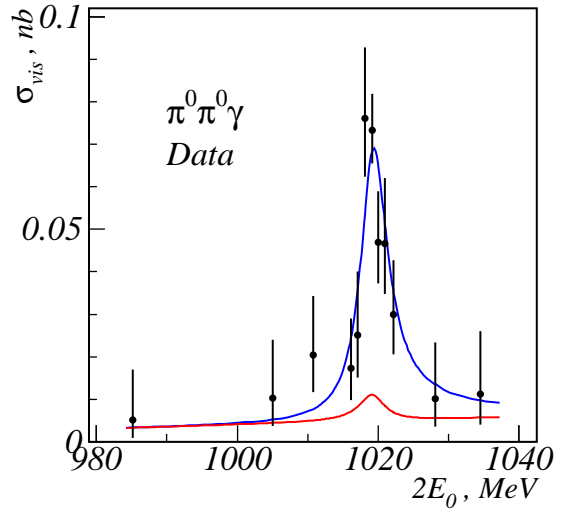}}
\vspace{1pt}
\caption{ Energy dependence of the visible
$e^+e^-\to\pi^0\pi^0\gamma$ cross section.
Points -- data, solid line -- fit, 
dotted line --  estimated background contribution from
$e^+e^-\to\omega\pi^0$ and
$\phi\to\eta\gamma$.}
\label{fig7}
\end{figure}

We would like to emphasize that all results presented in the previous section
are not based on model assumptions about $f_0$ structure.
The enhancement at large $m_{\pi\pi}$  (Fig.\ref{fig8})
is compatible only with large $f_0\gamma$ contribution. Further
analysis is carried out under assumption that the decay mechanism is a
pure $f_0\gamma$ and 
the mass spectrum can be described by the following
expression \cite{IVN2,IVN11}:
\begin{equation}
{ \frac{dBr(\phi\to\pi^o\pi^o\gamma)}{dm_{\pi\pi}}=
\frac{2m^2\Gamma(f_0\to\pi^0\pi^0)\Gamma(\phi\to f_0\gamma)}
{\pi{\left\vert D_f(m_{\pi\pi})\right\vert }^2},
 }
\label{SPEC}
\end{equation}
where $D_f(m_{\pi\pi})$ is an $f_0$ propagator, partial widths 
$\Gamma(f_0\to\pi^0\pi^0)$ and $\Gamma(\phi\to f_0\gamma)$ are functions
of $m_{\pi\pi}$. 
In the ``narrow resonance''  approximation \cite{NAR}
the results of the fit of the spectrum are the following: 
$$m_f = (984\pm12)~MeV,~\Gamma_f=(74\pm12)~MeV$$
in agreement with PDG data  \cite{PDG}. The coupling 
constant of $f_0$ with pions is found to be
$g^2_{f\pi^+\pi^-}/4\pi=(0.20\pm0.03)~GeV^2$.

On the other hand this simple approximation seems inadequate
for $f_0$ meson. In this case width corrections might be very large
\cite{NAR}. Instead of calculating such corrections we used
complete formulas from Refs.\cite{IVN2,IVN11} for the
``broad resonance'' fit. 
The formulas take into account final width of $f_0$
and strong coupling with $K\overline{K}$. 
The results of this fit are the following:
$$m_f = (971\pm6)~MeV,~\Gamma_f(m_f)=(188^{+48}_{-33})~MeV,$$
$$\frac{g^2_{fK^+K^-}}{4\pi}=2.10^{+0.88}_{-0.56} ~GeV^2,~
\frac{g^2_{f\pi^+\pi^-}}{4\pi}=0.51^{+0.13}_{-0.09} ~GeV^2,$$
$$g^2_{fK^+K^-}/g^2_{f\pi^+\pi^-}=4.1\pm0.9.$$
Comparison of the results of the fits shows strong model dependence
of $f_0$-meson parameters.
Both fitting curves in Fig.\ref{fig8} practically coincide,
although the ``broad resonance''
parameterization looks more realistic. In the latter case the
value of the coupling constant $g^2_{fKK}/4\pi$ obtained from the fit
agrees with the predictions of 4-quark MIT-bag
model  $(2.3~GeV^2$~\cite{IVN2,IVN13}) 
as well as the value of the branching ratio (\ref{EQA}). 
The corresponding predictions of the $K\overline{K}$ molecular
model \cite{IVN6,IVN11} and the 2-quark model \cite{IVN2} are
about 5 times lower. 

Of course, besides $f_0$ other intermediate states may contribute
to the decay under study, and most probably the
heavy and broad $\sigma$-state \cite{IVN11}.
This contribution can decrease values of $f_0$ coupling constants, 
but in any case such a state must produce
nearly flat $\pi^0\pi^0$ invariant mass spectrum,
not masking $f_0$-meson resonance signal in the mass region 
$m_{\pi\pi}>900~MeV.$ 
Theoretical estimations of branching ratios are also imprecise 
depending on a set of not well defined coupling
constants and may vary within a factor of two \cite{IVN11}. 
Thus, measured
$\phi\to f_0\gamma$ branching ratio, which is 5 times higher than
2-quark and $K\overline{K}$ model predictions
can be considered as another indication of 
significant 4-quark MIT-bag part in the $f_0$ meson not excluding
participation of other quark configurations.
To determine contributions of all possible
transition mechanisms and all quark configurations in $f_0$
one has to perform simultaneous analysis
of all $f_0$ data.

\section{Conclusion}

In the SND experiment  the
$\phi \to\pi^0\pi^0\gamma$ radiative decay
was observed for the first time 
and its branching ratio was measured.
Observed enhancement at high $m_{\pi\pi}$ in the invariant mass spectrum
together with angular distributions agreeing with scalar intermediate
$\pi^0\pi^0$ state show, that  $f_0(980)\gamma$ transition 
mechanism dominates in this decay.

Measured $f_0$ parameters are strongly model dependent due to
the fact that $f_0$ meson is broad and its mass is close to
$K\overline{K}$ threshold, making simple Breit-Wigner
description of the resonance inadequate.
 
Observed high decay  probability $\sim 10^{-4}$ 
gives evidence, that $f_0$ meson contains
significant 4-quark component but does not exclude participation of other
quark configurations.

\section{Acknowledgments}

Authors are grateful to N.N.Achasov for 
useful discussions and valuable comments.
The work is partially supported by RFBR 
(Grants No 96-02-19192; 96-15-96327; 97-02-18563) and
STP ``Integration'' (Grant No 274).

% now the references. delete or change fake bibitem. delete next three
%   lines and directly read in your .bbl file if you use bibtex.

\begin{table}
\caption{The experimental
mass spectrum for the $\phi\to\pi^o\pi^o\gamma$ decay
after background subtraction and acceptance corrections.
Only statistical error are shown.}
\vspace {2pt}
\label{IVN:T2}
\begin{tabular}{ll}
$m_{\pi\pi}~(MeV)$ & 
   $\frac{dBr(\phi\to\pi^o\pi^o\gamma)}{dm_{\pi\pi}}\cdot10^7~(MeV^{-1})$\\
\hline
600-620 & $0.61\pm0.84$ \\
620-640 & $1.26\pm0.89$ \\
640-660 & $0.59\pm0.68$ \\
660-680 & $1.66\pm0.84$ \\
680-700 & $1.62\pm0.78$ \\
700-720 & $0.81\pm0.61$ \\
720-740 & $2.86\pm0.94$ \\
740-760 & $1.63\pm0.76$ \\
760-780 & $1.73\pm0.80$ \\
780-800 & $2.69\pm0.96$ \\
800-820 & $2.50\pm0.94$ \\
820-840 & $2.25\pm0.90$ \\
840-860 & $4.58\pm1.21$ \\
860-880 & $2.32\pm0.91$ \\
880-900 & $3.80\pm1.11$ \\
900-920 & $3.59\pm1.14$ \\
920-940 & $3.74\pm1.17$ \\
940-950 & $8.41\pm2.47$ \\
950-960 & $8.93\pm2.62$ \\
960-970 & $6.50\pm2.34$ \\
970-980 & $3.74\pm1.93$ \\
980-990 & $6.20\pm2.57$ \\
990-1000& $1.60\pm1.60$ \\
\end{tabular}
\end{table}

\end{document}